\documentclass[11pt]{article}
\usepackage[margin=1in]{geometry}
\usepackage{times}

\usepackage{xcolor}
\usepackage[colorlinks=true,citecolor=blue,linkcolor=blue]{hyperref}

\hypersetup{
    colorlinks,
    linkcolor={red!50!black},
    citecolor={blue!50!black},
    urlcolor={blue!80!black}
}

\usepackage[]{amsmath,amssymb,amsfonts,latexsym,amsthm,enumerate,fullpage,xcolor,cite,bbm}
\usepackage[nameinlink, noabbrev, capitalize]{cleveref}
\usepackage{comment}
\usepackage{nicefrac}
\crefname{prop}{Proposition}{Propositions}
\crefname{ineq}{inequality}{inequalities}
\creflabelformat{ineq}{#2(#1)#3}

\newtheorem{theorem}{Theorem}
\newtheorem{lemma}{Lemma}

\newtheorem{fact}{Fact}

\newtheorem{corollary}{Corollary}
\newtheorem{definition}{Definition}

\crefname{THM}{Theorem}{Theorems}

\usepackage{subcaption}
\usepackage{graphicx}
\usepackage{tikz}
\usepackage{color, colortbl}
\definecolor{LightCyan}{rgb}{0.88,1,1}
\definecolor{Gray}{gray}{0.9}

\newcommand{\E}{\mathbb{E}}
\newcommand{\N}{\mathbb{N}}
\newcommand{\F}{\mathbb{F}}

\newcommand{\polylog}{\operatorname{polylog}}
\newcommand{\supp}{\text{support}}
\newcommand{\pr}{{\prime}}

\newcommand{\U}{\mathbf{U}}
\newcommand{\X}{\mathbf{X}}
\newcommand{\Y}{\mathbf{Y}}

\newcommand{\A}{\mathbf{A}}
\newcommand{\B}{\mathbf{B}}

\newcommand{\Z}{\mathbf{Z}}

\newcommand{\ttZ}{\tilde{\tilde{\mathbf{Z}}}}
\newcommand{\ttX}{\tilde{\tilde{\mathbf{X}}}}
\newcommand{\tY}{\tilde{\mathbf{Y}}}

\newcommand{\zo}{\{0,1\}}
\newcommand{\oz}{\{-1,1\}}

\newcommand{\Ext}{\mathsf{Ext}}

\newcommand{\nmExt}{\mathsf{nmExt}}

\newcommand{\NOF}{\mathsf{NOF}}

\newcommand{\cond}{\mathsf{2Cond}}

\newcommand{\eps}{\varepsilon}

\newcommand{\LRE}{\mathsf{LRE}}

\usepackage{subfiles}
\newcommand{\dobib}{
    \bibliographystyle{alpha}
    \bibliography{references} 
}

\begin{document}
\renewcommand{\dobib}{}

\title{Leakage-Resilient Extractors\\against Number-on-Forehead Protocols}

 \author{Eshan Chattopadhyay\thanks{Supported by a Sloan Research Fellowship and NSF CAREER Award 2045576.}\\ Cornell University\\ \texttt{eshan@cs.cornell.edu} \and Jesse Goodman\thanks{Supported by a Simons Investigator Award (\#409864, David Zuckerman).}\\ The University of Texas at Austin\\ \texttt{jpmgoodman@utexas.edu}}
\date{}

 \maketitle

\begin{abstract}

Given a sequence of $N$ independent sources $\mathbf{X}_1,\mathbf{X}_2,\dots,\mathbf{X}_N\sim\{0,1\}^n$, how many of them must be good (i.e., contain some min-entropy) in order to extract a uniformly random string? This question was first raised by Chattopadhyay, Goodman, Goyal and Li (STOC '20), motivated by applications in cryptography, distributed computing, and the unreliable nature of real-world sources of randomness. In their paper, they showed how to construct explicit low-error extractors for just $K \geq N^{1/2}$ good sources of polylogarithmic min-entropy. In a follow-up, Chattopadhyay and Goodman improved the number of good sources required to just $K \geq N^{0.01}$ (FOCS '21). In this paper, we finally achieve $K=3$.

Our key ingredient is a near-optimal explicit construction of a new pseudorandom primitive, called a leakage-resilient extractor (LRE) against number-on-forehead (NOF) protocols. Our LRE can be viewed as a significantly more robust version of Li's low-error three-source extractor (FOCS '15), and resolves an open question put forth by Kumar, Meka, and Sahai (FOCS '19) and Chattopadhyay, Goodman, Goyal, Kumar, Li, Meka, and Zuckerman (FOCS '20). Our LRE construction is based on a simple new connection we discover between multiparty communication complexity and non-malleable extractors, which shows that such extractors exhibit strong average-case lower bounds against NOF protocols.

\dobib

\end{abstract}

\maketitle

\section{Introduction}

Randomness is a surprisingly useful tool, with important applications in algorithm design, combinatorics, cryptography, distributed computing, and more \cite{vadhan2012pseudorandomness}. This raises the natural question:
\begin{center}
    Where does this randomness come from?
\end{center}
In the real world, random bits are harvested from a variety of natural phenomena, including atmospheric noise, thermal noise, radioactive decay, and other quantum phenomena \cite{herrero2017quantum}. Unfortunately, most applications of randomness require access to uniformly random bits, while the bits produced in nature can have various correlations and biases. This motivates the need for a device that can purify these weak sources of randomness into perfectly random bits. A \emph{randomness extractor} is exactly such a device.

\setcounter{definition}{-1} 
\begin{definition}[Randomness extractor]\label{def:extractor}
Let \(\mathcal{X}\) be a family of distributions over \(\zo^n\). A function \(\Ext:\zo^n\to\zo^m\) is called an \emph{extractor for \(\mathcal{X}\) with error \(\eps\)} if for every \(\X\in\mathcal{X}\),
    \[
    |\Ext(\X)-\U_m|\leq\eps,
    \]
    where \(\U_m\) denotes the uniform distribution over \(\zo^m\) and \(|\cdot|\) denotes statistical distance.
\end{definition}

Ever since the classical work of von Neumann \cite{von1951various}, extractors have become a central object in complexity theory, finding applications in cryptography \cite{bennett1988privacy}, combinatorics \cite{li2023two}, coding theory \cite{guruswami2004better}, and more. Along the way, a beautiful theory has grown around extractors, resulting in a rich literature spanning four decades and hundreds of papers. The fundamental question underlying much of this work is,
\begin{center}
    Which family of distributions \(\mathcal{X}\) should we try to extract from?
\end{center}

Of course, we would like the family \(\mathcal{X}\) to be as general as possible, but some assumptions are necessary in order for extraction to actually \emph{be} possible. First, each source \(\X\in\mathcal{X}\) must clearly have \emph{some} randomness in it. In the extractor literature, it is common to formalize this using the notion of \emph{min-entropy}, defined as
\[
H_\infty(\X):=\min_{x\in\supp(\X)}\log\left(1/\Pr[\X=x]\right).
\]
However, even this assumption is not enough, and a classic impossibility result \cite{chor1988unbiased} says that it is impossible to extract even from the family of distributions with nearly the \emph{maximum} amount of min-entropy \(k=n-1\). As such, researchers have looked towards additional assumptions to make on the family \(\mathcal{X}\).

\subsubsection*{Independent sources}
Perhaps the oldest and most well-studied assumption to make on the family \(\mathcal{X}\) is that each source \(\X\in\mathcal{X}\) actually consists of \emph{several} independent sources \(\X=(\X_1,\dots,\X_N)\), each over \(n\) bits, and each guaranteed to have some min-entropy, \(k\).\footnote{In this paper, \(N\) and \(n\) are completely unrelated variables (unlike many papers in the field that take \(N\) to mean \(2^n\)).} A systematic study of this model began with the seminal works of Vazirani \cite{vazirani1987strong} and Chor and Goldreich \cite{chor1988unbiased}, and has continued until the present day. Today, after a long line of work stretching over three decades (e.g., \cite{vazirani1987strong,chor1988unbiased,barak2006extracting,bourgain2005more,rao2009extractors,li2015three,chattopadhyay2019explicit,li2023two}), we now have near-optimal extractors for the independent source model.\footnote{In more detail, we now have explicit three-source extractors for \(k=\polylog(n)\) entropy and exponentially small error \cite{li2015three}, explicit two-source extractors for \(k=\polylog(n)\) entropy and polynomially small error \cite{chattopadhyay2019explicit}, and explicit two-source extractors for \(k=O(\log n)\) entropy and constant error \cite{li2023two}.}

\subsubsection*{Adversarial sources}

While the study of extractors for independent sources has seen great success, it is not always clear if this is actually a realistic model. In the real world, natural sources of randomness can be unreliable, and may sometimes (at unknown times) produce samples with no entropy at all. This motivates the study of a more general model, where each source need not have a min-entropy guarantee. Such a model was introduced by Chattopadhyay, Goodman, Goyal and Li \cite{ourAdversarial}, and are known as \emph{adversarial sources}.\footnote{In the original paper \cite{ourAdversarial}, the authors also consider a more general notion of adversarial sources, where each bad source can depend on up to \(d\) good sources. However, the majority of the work \cite{ourAdversarial}, and all of the follow-up work \cite{ourSmallSpace}, focuses on the fundamental setting of \(d=0\) (which, as we will see, already enjoys many applications). We take the same focus, here.}

\begin{definition}[Adversarial source]\label{def:adversarial-sources}
    A source \(\X\sim(\zo^n)^N\) is called an \emph{\((N,K,n,k)\)-adversarial source} if it is of the form \(\X=(\X_1,\X_2,\dots,\X_N)\), where each \(\X_i\) is an independent source over \(n\) bits, and at least \(K\) of them are \emph{good:} i.e., there is some set \(S\subseteq[N]\) of size \(|S|\geq K\) such that \(H_\infty(\X_i)\geq k\) for all \(i\in S\).
\end{definition}

If one could construct explicit extractors for adversarial sources, one could harvest uniform bits in a much more robust way, which doesn't completely break down whenever a source fails to output min-entropy. Moreover, as discussed in \cite{ourAdversarial}, extractors for adversarial sources have applications in cryptography (such as in generating \emph{common random strings} (CRS) in the presence of adversaries \cite{goyal2008universally,garg2011bringing,groth2014cryptography}) and distributed computing (i.e., in the execution of \emph{collective coin flipping protocols} \cite{ben1985collective,chor1985bit}).\footnote{CRS's are useful, because they enable the construction of otherwise impossible crytographic primitives, such as \emph{non-interactive zero-knowledge (NIZK) proofs} \cite{blum1988non} and \emph{universally composable (UC) commitment schemes} \cite{canetti2001universally}. And the ability to generate a CRS \emph{in the presence of adversaries} has natural applications in blockchains, where leader election protocols must succeed even if some players are malicious.} In these applications, it is often crucial to construct extractors that have negligible error \(\eps=n^{-\omega(1)}\) \cite{dodis2004possibility}. This leads us to our main motivating question:

\begin{center}
    Can we construct explicit low-error extractors for adversarial sources?
\end{center}

\subsubsection*{Prior constructions}

While the adversarial source model was first introduced in \cite{ourAdversarial}, it was implicitly studied in several prior works \cite{konig2004extracting,koenig2005generalized,goldwasser2005distributed,lee2005extracting,chattopadhyay2016extractors}. There, it was shown that a two-source extractor can be applied in a black-box manner to construct extractors for adversarial sources with \(K=2\) good sources. Plugging in the best-known low-error two-source extractors \cite{bourgain2005more,lewko2019explicit}, this gives low-error extractors for adversarial sources with \(K=2\) good sources of min-entropy roughly \(k\geq\frac{4}{9}n\). In related work \cite{bourgain2006estimates,kamp2011deterministic}, it was shown how to extract from \(K=O(1)\) good sources of min-entropy \(k=0.01n\).

Following these early papers, all subsequent work has focused on reducing the entropy requirement of the good sources (while using many more of them). In particular, Kamp, Rao, Vadhan and Zuckerman \cite{kamp2011deterministic} showed how to construct low-error extractors for adversarial sources with roughly \(K=\sqrt{N}\) good sources of min-entropy \(k\geq n^{0.99}\). Then, in \cite{ourAdversarial}, it was shown how to reduce the entropy requirement to just \(k\geq\polylog(n)\), while still using roughly \(K=\sqrt{N}\) good sources. Finally, a follow-up work \cite{ourSmallSpace} showed how to extract from just \(K=N^{0.01}\) good sources of min-entropy \(k\geq \polylog(n)\). To date, these remained the best-known low-error extractors for adversarial sources.

\subsection{Our results}

\subsubsection*{Extractors for adversarial sources}

In this paper, we construct explicit low-error extractors for adversarial sources with just \(K=3\) good sources, dramatically improving on the previous best requirement of \(K=N^{0.01}\). We prove the following.

\begin{theorem}[Extractors for adversarial sources]\label{thm:main:extractors-adversarial-sources}
There exist universal constants \(C,\gamma>0\) such that the following holds. There exists an explicit extractor \(\Ext:(\zo^n)^N\to\zo^m\) for \((N,K,n,k)\)-adversarial sources with \(K=3\) good sources of min-entropy \(k\geq\log^Cn\), which has output length \(m=\lceil k^\gamma\rceil\) and error \(\eps=2^{-\lceil k^\gamma\rceil}\).
\end{theorem}

To summarize, the original paper on adversarial sources required \(K=\sqrt{N}\) good sources \cite{ourAdversarial}, while the follow-up work \cite{ourSmallSpace} improved this requirement to \(K=N^{0.01}\). In this work, we reduce the number of good sources required to just \(K=3\), no matter how many sources, \(N\), there are in total (including super-constant and beyond). This essentially ``finishes off'' the problem of extracting from adversarial sources, as any further improvement is equivalent to constructing improved low-error two-source extractors -- one of the foremost remaining challenges in extractor theory.\footnote{As alluded to in the footnote preceding \cref{def:adversarial-sources}, there is still a more general notion of adversarial sources (where the bad sources can \emph{depend} on the good sources) for which much exciting work remains to be done.}

\cref{thm:main:extractors-adversarial-sources} has some interesting interpretations in the context of both extractors and cryptography. In the context of cryptography, this result gives a way for any number \(N\) of players to agree on a uniformly random string, provided that just \(3\) of them are honest. In the context of extractors, this result gives a way to extract from just \(3\) sources of weak randomness, hidden among any number of bad sources. In this way, it can be thought of as a significantly more robust version of Li's classical three-source extractor \cite{li2015three}.

Finally, our construction is significantly simpler than prior constructions of extractors for adversarial sources. Indeed, \cref{thm:main:extractors-adversarial-sources} follows immediately from a powerful new object that we construct, called a \emph{leakage-resilient extractor (LRE) against number-on-forehead (NOF) protocols}. Such objects lie directly at the intersection of extractor theory and communication complexity, and we discuss them next.

\subsubsection*{Leakage-resilient extractors against number-on-forehead protocols}

\emph{LREs against NOFs} are a brand new type of pseudorandom primitive, first introduced in \cite{kumar2019leakage} (under the name \emph{cylinder intersection extractors}). There are two equivalent ways to define LREs against NOFs: either as (1) a robust extractor for independent sources, or (2) a strong average-case lower bound against communication protocols. Let us take a moment to motivate and present definition (2). To do so, we must briefly recall a rich subfield of complexity theory, called \emph{(multiparty) communication complexity} \cite{yao1979some,dolev1992determinism,chandra1983multi}.

In multiparty communication complexity, the goal is to understand the amount of communication necessary to compute a function \(f:(\zo^n)^N\to\zo^m\) when its input is split up among \(N\) parties. In the most well-studied model, the parties must determine the value of \(f\) via a \emph{number-on-forehead} (NOF) protocol \cite{chandra1983multi} (see \cref{def:NOFs}). Here, each party \(i\in[N]\) receives one piece of the input \(x_i\in\zo^n\), which is metaphorically written on their forehead. Then, the parties take turns writing a bit on a public chalkboard, using all of the inputs \emph{except} the one written on their forehead. The protocol continues until everyone knows the value of \(f\), and the NOF communication complexity of \(f\) is the number of bits written on the chalkboard (in the worst-case over all possible inputs \(x\), for the best possible protocol).

As it turns out, NOF protocols are an incredibly powerful computational model, capable of simulating many other well-studied models. This means that explicit lower bounds against NOF protocols immediately yield explicit lower bounds against important models such as \(\mathsf{ACC}^0\) circuits \cite{yao1990acc,razborov1993nomega,beigel1994acc} and low-degree polynomials over \(\F_2\) \cite{viola2009guest}. Because of this, researchers have spent a significant amount of effort trying to construct \emph{explicit NOF lower bounds} (i.e., explicit functions that have high NOF communication complexity) \cite{babai1992multiparty}, in the hopes of achieving bounds that are strong enough to yield breakthrough lower bounds for these other computational models.

The most basic form of an explicit NOF lower bound that one might hope for is a \emph{worst-case bound}. This is an explicit function \(f\) such that for every NOF protocol \(\Pi\) computing \(f\), there is \emph{some} input \(x\) for which \(\Pi\) must write many bits on the blackboard in order to compute \(f(x)\).

An even stronger lower bound one might pursue is an \emph{average-case bound}. This is an explicit function \(f\) such that for every NOF protocol \(\Pi\) computing \(f\), it holds that for \emph{many} inputs \(x\), \(\Pi\) must write many bits \(\mu\) on the blackboard in order to compute \(f(x)\). Equivalently, for every protocol \(\Pi^\pr\) that writes \(<\mu\) bits on the blackboard (on every input), it holds that \(\Pi^\pr(\X)\) is highly uncorrelated with \(f(\X)\), where \(\X\sim(\zo^n)^N\) is a uniform random variable.

This leads us to the strongest of all lower bounds, which we might call a \emph{\(\mathcal{X}\)-average-case lower bound}. This is an explicit function \(f:(\zo^n)^N\to\zo^m\) such that for all protocols \(\Pi^\pr\) that write \(<\mu\) bits on the blackboard, and all sources \(\X\in\mathcal{X}\), it holds that \(\Pi^\pr(\X)\) is highly uncorrelated with \(f(\X)\).\footnote{The average-case lower bound definition corresponds to the case where \(\mathcal{X}\) consists of a single source, namely the uniform one.} Given that each \(\X\sim(\zo^n)^N\) consists of \(N\) parts \(\X=(\X_1,\dots,\X_N)\), it is natural to take \(\mathcal{X}\) to be the family of independent sources, where each part \(\X_i\sim\zo^n\) has min-entropy \(k\).

As it turns out, this is exactly the definition of an LRE against NOF protocols (see \cref{def:LREs-against-NOFs} for a formal definition). Indeed, in order for \(f\) to be highly uncorrelated with \(\Pi^\pr\), a trivial requirement is that \(f\) looks uniform on any \(\X\in\mathcal{X}\) (for otherwise it is correlated with a trivial protocol that always outputs a constant). Moreover, it is not too difficult to see that the output \(f(\X)\) must look uniform \emph{and remain uniform}, even conditioned on fixing the ``NOF leakage'' \(\Pi^\pr(\X)\). In this way, since \(\mathcal{X}\) denotes the family of independent sources, \(f\) can be thought of as a more robust type of extractor for independent sources -- definition (1).

LREs against NOFs were first introduced (under the name \emph{cylinder intersection extractors}) in the work \cite{kumar2019leakage}, and systematically studied in \cite{ourLREs}. It has long been known \cite{ourAdversarial} that if one could construct low-error LREs against NOFs for polylogarithmic min-entropy (and three parties), then one would immediately obtain \cref{thm:main:extractors-adversarial-sources}. However, such an object seemed challenging to construct, as it must not only be as strong as Li's classical three-source extractor \cite{li2015three}, but also offer powerful robustness guarantees. In this paper, we finally construct them.

\begin{theorem}[LREs against NOFs]\label{thm:main:LREs}
There exist universal constants \(C,\gamma>0\) such that the following holds. There exists an explicit three-source extractor \(\Ext:(\zo^n)^3\to\zo^m\) for min-entropy \(k\geq\log^Cn\), which is leakage-resilient against number-on-forehead protocols with \(\mu=\lceil k^\gamma\rceil\) bits of communication, and which has output length \(m=\lceil k^\gamma\rceil\) and error \(\eps=2^{-\lceil k^\gamma\rceil}\).
\end{theorem}

Prior to our work, all LREs against NOFs required high min-entropy, \(k\). In the paper that introduced these objects \cite{kumar2019leakage}, the authors showed how to construct such LREs for min-entropy \(k=0.99n\), by observing that existing average-case NOF lower bounds can handle a little missing entropy in the inputs for free. In the follow-up work \cite{ourLREs}, the authors improve this entropy requirement to \(k=0.3n\), via intricate exponential sum estimates. In this paper, we reduce the entropy requirement to \(k=\polylog(n)\), via a simple new reduction to \emph{non-malleable extractors} (in fact, a very weak version of this object - see \cref{def:weak-NMEs}).

Non-malleable extractors (NMEs) are a different, much more well-studied flavor of robust extractor, and were first introduced in \cite{dodis2009non,cheraghchi2017non}. We find the fact that there exists a reduction from LREs to NMEs to be quite surprising, for two reasons:
\begin{itemize}
    \item First, prior to this work, LREs and NMEs were thought to be incomparable. Indeed, even though they are both robust flavors of randomness extractors, they are robust in seemingly incomparable ways. In particular, while an LRE must defend against \emph{arbitrary} leaks, these leaks can only act on \(N-1\) of the inputs. An NME, on the other hand, must only defend against ``self-leaks'' (where the leaks are the NME itself, called with slightly modified inputs), but these leaks can depend on all \(N\) of the inputs. This work shows that NMEs are, in fact, the stronger object.
    \item Second, since LREs against NOFs are simply a strong type of lower bound against NOF protocols (as discussed above), and we show that every non-malleable extractor is such an LRE, our work also shows that \emph{non-malleable extractors witness strong NOF lower bounds}. In fact, the parameters we obtain show that NMEs achieve essentially the best-known lower bounds against NOF protocols \cite{babai1992multiparty}, and are even able to do so in the (stronger) low-entropy setting.\footnote{Indeed, while \cref{thm:main:LREs} is stated for just three parties, our full result (\cref{thm:technical:LREs}) works for any number \(N\) of parties with min-entropy \(k\), and achieves a lower bound of the form \(k^{\Omega(1)}/2^N\) for any \(k\geq\polylog(n)\). Since the current best lower bounds against NOF protocols are of the form \(\Omega(n/2^N)\), our result comes close to matching these bounds, and does so in the stronger ``low-entropy'' setting. (In the full min-entropy setting \(k=n\), one can obtain ``NME-based'' NOF lower bounds that essentially match the state-of-the-art -- for more details, see the full version.)} This provides a brand new family of black-box functions that are hard, and gives further support for a recent blossoming line of work on extractor-based lower bounds. In particular, prior to this work, it was known that basic extractors witness the best-known explicit lower bounds against general circuits \cite{LY22}, DNFs of parities \cite{CS16b}, linear ROBPs \cite{gryaznov2022linear,chattopadhyay2023hardness}, and any model that shrinks under random restrictions \cite{GG24}. In this paper, we show that essentially the best-known \emph{NOF lower bounds} are also directly witnessed by basic extractors - namely, non-malleable extractors.
\end{itemize}

Finally, we note that the reduction from LREs against NOFs to NMEs is simple. In particular, it just involves viewing (and slightly rephrasing) standard Cauchy-Schwarz-based proofs of NOF lower bounds in the language of extractors.\footnote{In more detail, we show that NMEs (for independent sources) have small \emph{cube norm}, which implies that they have large NOF communication complexity. A similar idea can be used to show that NMEs for affine sources (in fact, even just \emph{directional affine extractors} \cite{gryaznov2022linear}) have small \emph{Gowers norm}, and thus exhibit correlation bounds against low-degree \(\F_2\)-polynomials.} Thus, beyond this insight, all of the work in constructing our low-entropy LREs against NOFs (\cref{thm:main:LREs}) goes into constructing low-entropy NMEs, which is done via a composition of basic tools from extractor theory, combined with a slightly clever fixing argument.

\subsubsection*{Organization}

The rest of the paper is organized as follows. In \cref{sec:prelims}, we start with some basic preliminaries. Then, in \cref{sec:LREs-to-NMEs}, we give the reduction from LREs to NMEs. Then, we construct our explicit NMEs in \cref{sec:explicit-NMEs}. Finally, in \cref{sec:wrapping-up}, we instantiate the reduction with our explicit NMEs to obtain our explicit LREs (\cref{thm:main:LREs}), and recall the standard reduction from extractors for adversarial sources to LREs to immediately obtain \cref{thm:main:extractors-adversarial-sources}.

\dobib

\section{Preliminaries}\label{sec:prelims}

\paragraph{Notation}

First, let us introduce some helpful notation. To start, for an integer \(n\in\N\), we let \([n]:=\{1,2,\dots,n\}\). Given a string \(x\in\zo^n\), we let \(x_i\) denote the value it holds at its \(i^\text{th}\) coordinate, and for any subset \(S\subseteq[n]\), we let \((x_i)_{i\in S}\) denote the concatenation of all bits \(x_i\) in increasing order of \(i\). Speaking of concatenation, we sometimes represent it using the symbol \(\circ\), in the sense that \(x\circ y:=(x,y)\). In particular, throughout this paper, \(\circ\) does \emph{not} denote function composition. Another perhaps unconventional notation we use throughout is that \(N,n\) will always denote \emph{completely unrelated} integers. Namely, we \emph{do not} take \(N\) to mean \(2^n\). (The same goes for \(K\) and \(k\).)

Now, given a subset \(S\subseteq[n]\), we write \(x_S\) as shorthand for \((x_i)_{i\in S}\), and for any \(m\in[n]\), we write \(x_{\leq m}\) as shorthand for \(x_{[m]}\) (i.e., the first \(m\) bits of \(x\)), and define \(x_{<m}\) and \(x_{>m}\) and \(x_{\geq m}\) in the analogous way. Next, for any \(i\in[n]\) we define \(x_{-i}\) to mean \(x_{[n]\setminus\{i\}}\). All of this notation extends naturally to larger alphabets. For example, for any alphabet \(\Sigma\) and \(x\in\Sigma^n\), we let \(x_i\) denote the \(i^\text{th}\) symbol in \(x\). If \(\Sigma=\zo^m\), then \(x_i\) denotes the \(i^\text{th}\) consecutive chunk of \(m\) bits. Finally, for any positive integer \(t\), we let \(\F_{2^t}\) denote the finite field over \(2^t\) elements, and whenever we write \(x\cdot y\) for \(x,y\in\zo^t\), we intend this to mean their product over \(\F_{2^t}\). As usual, all logs will be base \(2\).

\subsection{Probability}

Throughout, we use bold font, such as \(\X\), to denote random variables. The \emph{support} of \(\X\), denoted \(\supp(\X)\), contains all elements \(x\) such that \(\Pr[\X=x]>0\), and we write \(\X\sim V\) if \(\supp(\X)\subseteq V\). As per tradition in the extractor literature, we will often refer to random variables as \emph{sources}. Furthermore, for any set \(V\) and \(S\subseteq V\), we let \(\U_S\) denote a uniform random variable over \(S\), and use the shorthand \(\U_m\) when \(S=\zo^m\). That is, \(\U_m\) is a uniform random variable over bitstrings of length \(m\).

Next, when talking about extractors, the canonical distance that is used is called the \emph{statistical distance}:

\begin{definition}[Statistical distance]
The \emph{statistical distance} between two random variables \(\X,\Y\sim V\) is
\begin{align*}
|\X-\Y|&:=\max_{S\subseteq V}|\Pr[\X\in S]-\Pr[\Y\in S]|=\frac{1}{2}\sum_{v\in V}|\Pr[\X=v]-\Pr[\Y=v]|.
\end{align*}
If \(|\X-\Y|\leq\eps\), we write \(\X\approx_\eps\Y\) and say \(\X,\Y\) are \emph{\(\eps\)-close}. Otherwise, we say they are \emph{\(\eps\)-far}. Moreover, if \(|\X-\Y|=0\), we write \(\X\equiv\Y\).
\end{definition}

Next, we introduce three extremely useful facts about statistical distance. First, when dealing with extractors, the classical goal is to show that its output is close to uniform, \(\U_m\). Towards this end, the following tool is invaluable. It says that two random variables cannot get further apart if you process them with the same deterministic function.

\begin{fact}[Data-processing inequality]\label{lem:data-processing-inequality}
For any random variables \(\X,\Y\sim V\), and any function \(f:V\to W\),
\[
|\X-\Y|\geq|f(\X)-f(\Y)|.
\]
\end{fact}

Second, it is well-known that the \emph{triangle inequality} also holds for statistical distance:

\begin{fact}[Triangle inequality]\label{fact:triangle-inequality}
For any random variables \(\X,\Y,\Z\),
\[
|\X-\Y|\leq|\X-\Z|+|\Z-\Y|.
\]
\end{fact}

Third, the following fact is very useful when thinking about ``fixing'' random variables.

\begin{fact}[Averaging principle]\label{fact:averaging-principle}
    For any random variables \(\X,\Y\sim V\) and \(\Z,\Z^\pr\sim Z\) such that \(\Z\equiv\Z^\pr\),
    \[
    |\X\circ\Z-\Y\circ\Z^\pr|=\E_{z\sim\Z}[|(\X\mid\Z=z)-(\Y\mid\Z^\pr=z)|].
    \]
\end{fact}

Speaking of fixing random variables, the following standard lemma will be of great help whenever we want to argue that fixing one random variable doesn't cause another to lose too much min-entropy.

\begin{lemma}[Min-entropy chain rule \cite{maurer1997privacy}]\label{lem:chain-rule}
For any random variables \(\X\sim X,\Y\sim Y\) and any \(\eps>0\),
\[
\Pr_{y\sim\Y}\left[H_\infty(\X\mid\Y=y)\geq H_\infty(\X)-\log|Y|-\log(1/\eps)\right]\geq1-\eps.
\]
\end{lemma}

Finally, we record one last lemma, which (to the best of our knowledge) is new. The goal of this lemma is to formally ``capture'' (or describe) the randomness that is lost after applying a deterministic function \(f\) to a random variable \(\X\). It strengthens and simplifies \cite[Lemma 7]{ourLREs}.

\begin{lemma}[Dependency reversal]\label{lem:dependency-reversal}
For any random variable \(\X\sim X\) and deterministic function \(f:X\to Y\), there exists an independent random variable \(\A\sim A\) and deterministic function \(g:Y\times A\to X\) such that
\[
g(f(\X),\A)\equiv\X.
\]
\end{lemma}
\begin{proof}
We define an independent random variable \(\A\sim A:=X^Y\) as a sequence of independent random variables \(\A=(\A_y)_{y\in Y}\) where each \(\A_y\sim X\) is defined as
\[
\Pr[\A_y=x]:=\Pr[\X=x\mid f(\X)=y]\text{ for all }x\in X.
\]
If we then define the deterministic function \(g:Y\times A\to V\) as
\[
g(y,a):=a_y,
\]
it directly follows that \(g(f(\X),\A)\equiv\X\) via the law of total probability, as desired.
\end{proof}

\subsection{Randomness condensers}

We now recall the notion of \emph{condensers}, which are slightly weaker versions of extractors. While an extractor guarantees that its output is close to \emph{uniform} \(\U_m\), a condenser just guarantees that its output is close to \emph{having high min-entropy}. Formally, (two-source) condensers are defined as follows.

\begin{definition}[Two-source condensers]\label{def:condenser}
A function \(\cond:\zo^n\times\zo^n\to\zo^m\) is called an \emph{\((n,k)\to_\eps(m,\ell)\) two-source condenser}\footnote{Or equivalently, a two-source condenser with input entropy \(k\), output entropy \(\ell\), output length \(m\), and error \(\eps\).} if for any two independent sources \(\X,\Y\sim\zo^n\) each with min-entropy at least \(k\), \(\cond(\X,\Y)\) is \(\eps\)-close to some source \(\Z\sim\zo^m\) with min-entropy at least \(\ell\). We say \(\cond\) is \emph{strong} if
\begin{align*}
&\Pr_{y\sim\Y}[\cond(\X,y)\text{ is not \(\eps\)-close to a source with min-entropy at least \(\ell\)}]\leq\eps
\end{align*}
and
\begin{align*}
&\Pr_{x\sim\X}[\cond(x,\Y)\text{ is not \(\eps\)-close to a source with min-entropy at least \(\ell\)}]\leq\eps.
\end{align*}
\end{definition}

As it turns out, every two-source condenser is also a \emph{strong} two-source condenser, provided that we feed it two sources with a little more entropy. In particular, the following is true.

\begin{fact}[Every two-source condenser is strong]\label{fact:every-condenser-is-strong} For any \(n\geq k\) and \(m\geq \ell\), if \(\cond:\zo^n\times\zo^n\to\zo^m\) is an \((n,k)\to_\eps(m,\ell)\) two-source condenser, then it is a \emph{strong} \((n,k+m+\log(1/\eps))\to_\eps(m,\ell)\) two-source condenser.
\end{fact}

The proof of this fact is almost identical to a proof, due to Barak, that every two-source \emph{extractor} is also a strong two-source extractor (with slightly weaker parameters) \cite{rao2007exposition}. For completeness, we include its proof below. Beyond the ideas in Barak's analogous proof for two-source extractors, we need the following fact, which gives a nice characterization of sources that are close to having high min-entropy. (See \cite[Lemma 2]{goodman2024improved} for a proof.)

\begin{fact}[An alternative characterization for being close to high min-entropy \protect{\cite[Lemma 2.2]{zuckerman2007linear}}]\label{fact:condenser-equivalence}
For any \(n\geq k\), a source \(\X\sim\zo^n\) is \(\eps\)-close to a source of min-entropy at least \(k\) iff for every \(S\subseteq \zo^n\),
    \[
    \Pr[\X\in S]\leq|S|\cdot2^{-k}+\eps.
    \]
\end{fact}

With this characterization in hand, it is now straightforward to extend Barak's argument about two-source extractors \cite{rao2007exposition} to also work for two-source condensers, and thereby prove \cref{fact:every-condenser-is-strong}.

\begin{proof}[Proof of \cref{fact:every-condenser-is-strong}]
    We prove that for any independent sources \(\X,\Y\sim\zo^n\) with min-entropy at least \(k^\pr\),
    \begin{align*}
    &\Pr_{y\sim\Y}[\cond(\X,y)\text{ is not \(\eps\)-close to a source with min-entropy at least \(\ell\)}]\leq2^{k+m-k^\pr}
    \end{align*}
    and
    \begin{align*}
    &\Pr_{x\sim\X}[\cond(x,\Y)\text{ is not \(\eps\)-close to a source with min-entropy at least \(\ell\)}]\leq2^{k+m-k^\pr},
    \end{align*}
    and the result will follow by setting \(k^\pr=k+m+\log(1/\eps)\). Actually, we will just prove the first claim, as the proof of the second claim is identical. Towards this end, for any \(S\subseteq\zo^m\), define
    \[
    \mathsf{Bad}_S:=\{y\in\zo^m : \Pr[\cond(\X,y)\in S]>|S|\cdot2^{-\ell}+\eps\},
    \]
    and notice that \cref{fact:condenser-equivalence} tells us that \(\cond(\X,\U_{\mathsf{Bad}_S})\) is not \(\eps\)-close to a source with min-entropy at least \(\ell\). We must therefore have \(|\mathsf{Bad}_S|\leq 2^k\), or else \(\cond\) is not the condenser it claimed to be. Thus, if we define \(\mathsf{Bad}:=\bigcup_S\mathsf{Bad}_S\), we know that \(|\mathsf{Bad}|\leq2^{k+m}\). Now, again using \cref{fact:condenser-equivalence}, we have
    \begin{align*}
    &\Pr_{y\sim\Y}[\cond(\X,y)\text{ is not \(\eps\)-close to an \((m,\ell)\) source}]\\
    = &\Pr_{y\sim \Y}[\exists S\subseteq\zo^m : \Pr[\cond(\X,y)\in S] > |S|\cdot2^{-\ell}+\eps]\\
    = &\Pr_{y\sim\Y}[y\in\mathsf{Bad}]\leq|\mathsf{Bad}|\cdot2^{-k^\pr}=2^{k+m-k^\pr},
    \end{align*}
    as desired.
\end{proof}

\subsection{Communication complexity}

Finally, we provide a formal definition of \emph{number-on-forehead} (NOF) multiparty communication protocols.

\begin{definition}[Number-on-forehead protocol]\label{def:NOFs}
    A function \(\Pi:(\zo^n)^N\to\zo^\mu\) is called a \emph{number-on-forehead protocol} if for every \(i\in[\mu]\) there exist functions \(S_i:\zo^{i-1}\to[N]\) and \(g_i:\zo^{i-1}\times(\zo^n)^{N-1}\to\zo\) such that
    \[
    \Pi_i(x) = g_i(\Pi_1(x),\dots,\Pi_{i-1}(x),x_{-S_i(\Pi_1(x),\dots,\Pi_{i-1}(x))}).
    \]
    We say that \(\Pi\) is \emph{non-adaptive} if each \(g_i\) does not depend on its first \(i-1\) inputs and each \(S_i\) is a constant.
    \end{definition}

We say that a function \(f:(\zo^n)^N\to\zo^m\) has \(\eps\)-average-case NOF communication complexity \(>\mu\) if for any NOF protocol \(\Pi\) with output length \(\mu\), it holds that \(f(\X)\circ\Pi(\X)\approx_\eps\U\circ\Pi(\X)\), where \(\X,\U\) are independent uniform random variables. Equivalently, we say that \(f\) is \emph{\(\eps\)-hard} for such protocols.

To conclude this section, we record the following lemma, which says that an average-case NOF-hard function is still hard even if the input distributions are missing a little bit of min-entropy.

\begin{lemma}[NOF-hard functions can tolerate a little missing min-entropy \protect{\cite[Lemma 3]{ourLREs}}]\label{lem:NOF-with-missing-entropy}

Let \(\NOF:(\zo^n)^N\to\zo^m\) be a function with \(\eps\)-average-case NOF communication complexity \(>\mu\). Then for any independent sources \(\X_1,\dots,\X_N\sim\zo^n\) each with min-entropy at least \(k\), and any number-on-forehead protocol \(\Pi:(\zo^n)^N\to\zo^{\mu-2}\), it holds that
\begin{align*}
|\NOF(\X_1,\dots,\X_N)\circ\Pi(\X_1,\dots,\X_N)-\U_m\circ\Pi(\X_1,\dots,\X_N)|\leq\eps\cdot2^{N(n-k)}.
\end{align*}
\end{lemma}
\noindent As we will soon see (via \cref{def:LREs-against-NOFs}), another way to summarize this lemma is that any (average-case) NOF-hard function is automatically a basic LRE against NOF protocols, which can handle slightly less than full min-entropy.

\dobib

\section{A reduction from leakage-resilient extractors to non-malleable extractors}\label{sec:LREs-to-NMEs}

In order to construct our extractors for adversarial sources with \(K=3\) good sources (\cref{thm:main:extractors-adversarial-sources}), we will build a three-source leakage-resilient extractor (LRE) against number-on-forehead protocols (\cref{thm:main:LREs}), and the adversarial source extractor will easily follow (simply by calling the LRE on all triples of sources and XORing the results). To get started, we formally introduce LREs against NOFs.

\begin{definition}[LREs against NOFs]\label{def:LREs-against-NOFs}
    A function \(\LRE:(\zo^n)^N\) \(\to\zo^m\) is called a \emph{leakage-resilient extractor} for min-entropy \(k\) and error \(\eps\) against number-on-forehead protocols with \(\mu\) bits of communication if the following holds. For any independent sources \( \X_1,\dots,\X_N\sim\zo^n\) each with min-entropy at least \(k\), and any number-on-forehead protocol \(\Pi:(\zo^n)^N\to\zo^\mu\),
    \[
    \LRE(\X_1,\dots,\X_N)\circ\Pi(\X_1,\dots,\X_N)\approx_\eps\U_m\circ\Pi(\X_1,\dots,\X_N).
    \]
\end{definition}

LREs against NOFs were first introduced by Kumar, Meka and Sahai (under the name \emph{cylinder intersection extractors}) \cite{kumar2019leakage}, and systematically studied for the first time in \cite{ourLREs} (where the more general notion of LREs was introduced). Looking at \cref{def:LREs-against-NOFs}, it is easy to see that these are powerful objects, which simultaneously generalize both (1) extractors for independent sources and (2) average-case lower bounds against NOF protocols.\footnote{Indeed, setting \(\mu=0\) recovers object (1), while setting \(k=n\) recovers object (2).} In this section, we take our first step towards constructing them.

In order to build our LREs against NOFs, we provide a simple reduction from these objects to a new, weak type of non-malleable extractor, which we define below. Then, we will show how to build one.

\begin{definition}[Weak NMEs]\label{def:weak-NMEs}
    A function \(\nmExt:(\zo^n)^N\to\zo^m\) is called a \emph{weak non-malleable extractor} for min-entropy \(k\) and error \(\eps\) if the following holds. For any independent sources \(\X_1^0,\dots,\X_N^0,\X_1^1,\dots,\X_N^1\sim\zo^n\), each with min-entropy at least \(k\), and \(\X_i^0\equiv\X_i^1\) for all \(i\in[N]\),
\begin{align*}
\nmExt(\X_1^0,\dots,\X_N^0)\circ\left(\nmExt(\X_1^{b_1},\dots,\X_N^{b_N})\right)_{b\neq\vec{0}\in\zo^N}\approx_\eps\U_m\circ\left(\nmExt(\X_1^{b_1},\dots,\X_N^{b_N})\right)_{b\neq\vec{0}\in\zo^N}.
\end{align*}
\end{definition}

With these definitions in hand, we proceed with the first main lemma of this paper, which shows that weak NMEs are, in fact, LREs against NOFs.

\begin{lemma}[Weak NMEs \(\implies\) LREs against NOFs]\label{lem:main:reduction}
Let \(\nmExt:(\zo^n)^N\to\zo^m\) be a weak non-malleable extractor for min-entropy \(k\) and error \(\eps\). Then \(\nmExt\) is a leakage-resilient extractor against number-on-forehead protocols with \(\mu\) bits of communication for min-entropy \(k\) and error \(2^{m+\mu}\cdot(2\eps)^{1/2^N}\).
\end{lemma}

\begin{proof}
The proof, in fact, is not too difficult. It follows  from observing that the standard technique for obtaining NOF lower bounds (i.e., via the cube norm) both (1) automatically work for low entropy, and (2) can be interpreted as seeking out the non-malleability property put forth by \cref{def:weak-NMEs}. Indeed, the novelty of the  proof is just a shift in perspective.

To make things more formal, let \(\nmExt:(\zo^n)^N\to\zo^m\) be the weak non-malleable extractor from the lemma statement. We want to upper bound the statistical distance
\[
|\nmExt(\X)\circ\Pi(\X)-\U_m\circ\Pi(\X)|,
\]
where \(\Pi\) is an NOF protocol of length \(\mu\), and \(\X\) consists of \(N\) independent sources \(\X=(\X_1,\X_2,\dots,\X_N)\), each over \(n\) bits and each with min-entropy at least \(k\). The first issue is that lower-bounding the above expression amounts to achieving ``multi-bit'' hardness, whereas standard communication complexity seems to focus on proving hardness for Boolean-valued functions. However, there are standard tools from cryptography and extractor theory that can be exploited to get such a result. Here, we will use a ``non-uniform'' XOR lemma of Dodis, Li, Wooley, and Zuckerman \cite[Lemma 3.8]{dodis2014privacy}, which says that we can upper bound the above statistical distance by \(2^{m+\mu}\alpha\), provided we can show that
\[
\Bigg|\Big(\bigoplus_{i\in S}\nmExt_i(\X)\oplus\bigoplus_{j\in T}\Pi_j(\X)\Big)-\U_1\Bigg|\leq\alpha,
\]
for all \(S\neq\emptyset\subseteq[m],T\subseteq[\mu]\). Fix one such choice of \(S,T\), and define \(f(x):=\bigoplus_{i\in S}\nmExt_i(x)\) and \(g(x):=\bigoplus_{j\in T}\Pi_j(x)\). By the data-processing inequality (\cref{lem:data-processing-inequality}), the above statistical distance can be upper bounded by
\[
|f(\X)\circ g(\X)-\U_1\circ g(\X)|,
\]
where \(\U_1\) is independent of everything else. Applying the data-processing inequality once more, the above can further be upper bounded by
\[
|f(\X)\circ \Pi(\X)-\U_1\circ \Pi(\X)|.
\]
The goal is now to upper bound the above, by showing that if \(\nmExt\) started out as a weak non-malleable extractor, then \(f(\X)\) has low correlation with all NOF protocols. To show this, we can simply repeat known arguments originating in \cite{babai1992multiparty}\footnote{Our argument will follow along with the one that appears in the excellent lecture notes of Lovett \cite{lovett2019cse}.} through the lens of extractors.

To get started, note that
\begin{align*}
2|f(\X)\circ\Pi(\X)-\U_1\circ\Pi(\X)|&=2\E_{\pi\sim\Pi(\X)}|(f(\X)\mid\Pi(\X)=\pi)-\U_1|\\
&=\E_{\pi\sim\Pi(\X)}|\Pr[f(\X)=1\mid\Pi(\X)=\pi]-\Pr[f(\X)=0\mid\Pi(\X)=\pi]|\\
&\leq2^\mu\max_\pi|\Pr[f(\X)=1\land\Pi(\X)=\pi]-\Pr[f(\X)=0\land\Pi(\X)=\pi]|\\
&=2^\mu\cdot|\Pr[f(\X)=1\land\Pi(\X)=\pi]-\Pr[f(\X)=0\land\Pi(\X)=\pi]|,
\end{align*}
where the last line just fixes the worst transcript \(\pi\). Now, define the function \(f^\pr:=(-1)^f\) which swaps the output domain of \(f\) from \(\zo\) to \(\oz\), and define the function \(C:(\zo^n)^N\to\zo\) as the indicator of \(\Pi(x)=\pi\). Namely, \(C(x)=1\) iff \(\Pi(x)=\pi\). Using this, the most recent equation above is
\begin{align*}
    =2^\mu\cdot|\E[f^\pr(\X)\cdot C(\X)]|.
\end{align*}

We now focus on upper bounding \(|\E[f^\pr(\X)\cdot C(\X)]|\). This part of the argument is now standard, and we follow the exposition of Lovett \cite{lovett2019cse}. Towards this end, it is well-known (and not too hard to show) that \(C\) is the indicator of a ``cylinder intersection'' \cite{babai1992multiparty}. This means that there exist indicator functions \(C_1,\dots,C_N:(\zo^n)^{N-1}\to\zo\) so that \(C(x)=C_1(x_{-1})\cdot C_2(x_{-2})\cdot \dotsm \cdot C_N(x_{-N})\). Thus,
\begin{align*}
    |\E[f^\pr(\X)\cdot C(\X)]|=|\E[f^\pr(\X)\cdot C_1(\X_{-1})\cdot C_2(\X_{-2})\cdot \dotsm \cdot C_{N}(\X_{-N})]|
\end{align*}

Now, the main idea introduced in \cite{babai1992multiparty} is to use \(N\) iterations of the Cauchy-Schwarz inequality (for random variables) to get rid of the ``NOF leaks'' \(C_1,\dots,C_N\), and replace them with, in our language, ``weak non-malleable leaks.'' In more detail, recall that the Cauchy-Schwarz inequality tells us that for any real-valued random variables \(\A,\B\) it holds that \(|\E[\A\cdot\B]|\leq\left(\E[\A^2]\cdot\E[\B^2]\right)^{1/2}\). Applying this once, we can continue as follows.
\begin{align*}
    &=\left|\E_{\X_{-1}}\biggl[C_1(\X_{-1})\cdot\E_{\X_1}[f^\pr(\X)\cdot C_2(\X_{-2})\cdot\dotsm\cdot C_N(\X_{-N})]\biggl]\right|\\
    &\leq\Bigg(\E_{\X_{-1}}\left[\biggl(C_1(\X_{-1})\biggl)^2\right]\cdot\E_{\X_{-1}}\left[\biggl(\E_{\X_1}[f^\pr(\X)\cdot C_2(\X_{-2})\cdot\dotsm\cdot C_N(\X_{-N})]\biggl)^2\right]\Bigg)^{1/2}\\
    &\leq \Bigg(1\cdot\E_{\X_{-1}}\biggl[\E_{\X_1,\X_1^\pr}[f^\pr(\X)\cdot C_2(\X_{-2})\dotsm C_N(\X_{-N}) \cdot f^\pr(\X_{-1},\X_1^\pr)\cdot C_2(\X_{-1,-2},\X_1^\pr)\dotsm C_N(\X_{-1,-N},\X_1^\pr)]\biggl]\Bigg)^{1/2}\\
    &=\Bigg(\E_{\X,\X_1^\pr}\biggl[f^\pr(\X)\cdot C_2(\X_{-2})\dotsm C_N(\X_{-N}) \cdot f^\pr(\X_{-1},\X_1^\pr)\cdot C_2(\X_{-1,-2},\X_1^\pr)\dotsm C_N(\X_{-1,-N},\X_1^\pr)\biggl]\Bigg)^{1/2}.
\end{align*}

The first inequality is an application of the Cauchy-Schwarz inequality, where the random variable \(\B\) looks like an expectation. The second inequality is because the range of \(C_1\) is \(\zo\) (which, in particular, is at most \(1\) when squared), and because \(\X_1,\X_1^\pr\) are independent (and identically distributed). The last equality is immediate.

Notice that by using the Cauchy-Schwarz inequality once, we were able to ``remove'' or ``condition away'' one NOF leak \(C_1(\X_{-1})\), and effectively replace it with a ``non-malleable leak'' \(f^\pr(\X_{-1},\X_1^\pr)\) (i.e., instead of an arbitrary function leaking on some of the inputs, we turn it into the specific leak \(f^\pr\), acting on all inputs, or at least independent copies of them). To get the Cauchy-Schwarz inequality to work (to remove this NOF leak), we pulled out an expectation over the random variables on which it depends \(\X_{-1}\), leaving just \(\X_1\) in the internal expectation. When we execute it again, we'll once again pull out almost all random variables, this time leaving just \(\X_2\) in the internal expectation. This will let us get rid of the next NOF leak, \(C_2(\X_{-2})\). Thus, pulling out \(\X_{-2},\X_1^\pr\) and leaving in \(\X_2\), we can follow the argument above once again to obtain
\begin{align*}
\leq\biggl(\E_{\X,\X_1^\pr,\X_2^\pr}\biggl[&f^\pr(\X)\cdot C_3(\X_{-3})\dotsm C_N(\X_{-N})\cdot\\ &f^\pr(\X_{-1},\X_1^\pr)\cdot C_3(\X_{-3,-1},\X_1^\pr)\dotsm C_N(\X_{-N,-1},\X_1^\pr)\cdot\\
    &f^\pr(\X_{-2},\X_2^\pr)\cdot C_3(\X_{-3,-2},\X_2^\pr)\dotsm C_N(\X_{-N,-2},\X_2^\pr)\cdot\\
    &f^\pr(\X_{-1,-2},\X_1^\pr,\X_2^\pr)\cdot C_3(\X_{-3,-2,-1},\X_2^\pr,\X_1^\pr)\cdot\dotsm C_N(\X_{-N,-2,-1},\X_2^\pr,\X_1^\pr)
    \biggl]\biggl)^{1/2^2}.
\end{align*}
As the notation is getting cumbersome, it will be convenient to switch to something different. In particular, we define random variables \(\X_1^0,\X_2^0,\dots,\X_N^0,\X_1^1,\X_2^1,\dots,\X_N^1\) that are all independent and such that \(\X_i\equiv\X_i^0\equiv\X_i^1\). Then, for a fixed string \(b\in\zo^N\), we let \(\X^b\) denote \((\X_1^{b_1},\dots,\X_N^{b_N})\). With this notation, the above expression is much easier to write:
\[
=\E\left[\prod_{\substack{b\in\zo^N\\:b_i=0\forall i>2}}f^\pr(\X^b)\cdot C_3(\X_{-3}^b)\cdot C_4(\X_{-4}^b)\dotsm C_N(\X_{-N}^b)\right]^{1/4}
\]
Finally, repeating the argument with the remaining leaks \(\{C_i\}_{i\geq3}\) immediately gives

\begin{align*}
&\leq\left(\E\left[\prod_{b\in\zo^N}f^\pr(\X^b)\right]\right)^{1/2^N}\\
&=\left(\Pr\left[\prod_{b\in\zo^N}f^\pr(\X^b)=1\right]-\Pr\left[\prod_{b\in\zo^N}f^\pr(\X^b)=-1\right]\right)^{1/2^N}\\
&=\left(\Pr\left[\bigoplus_{b\in\zo^N}f(\X^b)=0\right]-\Pr\left[\bigoplus_{b\in\zo^N}f(\X^b)=1\right]\right)^{1/2^N}\\
&=\left(2\left|\bigoplus_{b\in\zo^N}f(\X^b)-\U_1\right|\right)^{1/2^N}\\
&\leq\left(2\cdot\Big|f(\X)\circ(f(\X^b))_{b\neq0\in\zo^N}-\U_1\circ(f(\X^b))_{b\neq0\in\zo^N}\Big|\right)^{1/2^N}
\end{align*}
where the last line is a simple application of the data-processing inequality. This almost looks like the weak non-malleable extractor definition, but recall \(f\) itself is of the form \(f(x):=\oplus_{i\in S}\nmExt_i(x)\) for some nonempty \(S\). Since this is just an XOR of some output bits of \(\nmExt\), it is clearly a deterministic function of it, and thus we may apply the data-processing inequality (\cref{lem:data-processing-inequality}) once more to upper bound the above by
\begin{align*}
\leq\Bigg(2\cdot\Big|\nmExt(\X)\circ(\nmExt(\X^b))_{b\neq0\in\zo^N}
-\U_m\circ(\nmExt(\X^b))_{b\neq0\in\zo^N}\Big|\Bigg)^{1/2^N},
\end{align*}
and since we were given that \(\nmExt\) is a weak-non-malleable extractor for min-entropy \(k\) and error \(\eps\), it follows that this is
\[
\leq(2\eps)^{1/2^N}.
\]
To summarize, we have finished showing that
\[
\Bigg|\Big(\bigoplus_{i\in S}\nmExt_i(\X)\oplus\bigoplus_{j\in T}\Pi_j(\X)\Big)-\U_1\Bigg|\leq(2\eps)^{1/2^N}.
\]
As discussed at the beginning of the proof, the non-uniform XOR lemma allows us to therefore conclude
\[
\left|\nmExt(\X)\circ\Pi(\X)-\U_m\circ\Pi(\X)\right|\leq2^{m+\mu}\cdot(2\eps)^{1/2^N},
\]
as desired.
\end{proof}

\dobib

\section{An explicit construction of weak non-malleable extractors}\label{sec:explicit-NMEs}

Now that we know that weak NMEs automatically give LREs against NOFs, we turn our attention to explicitly constructing them. We will prove the following lemma.

\begin{lemma}[Explicit weak NMEs]\label{lem:main:explicit-wNMEs}
There exist universal constants \(C,\gamma>0\) such that the following holds. There exists an explicit weak non-malleable extractor \(\nmExt:(\zo^n)^N \to\zo^m\) for min-entropy \(k\geq\log^Cn\), which has output length \(m=\lceil k^\gamma/2^N \rceil\) and error \(\eps=2^{-\lceil k^\gamma/2^N\rceil}\).
\end{lemma}

In order to prove this lemma, we show a simple way to construct weak NMEs from basic, black-box tools. As it turns out, one of these tools is \emph{itself} a function that is hard against number-on-forehead protocols! Thus, we will have traveled from LREs against NOFs, to weak NMEs, and back to functions that are hard against NOF protocols (but for one less party). As a result, we ultimately provide a way to convert a black-box function that is hard for NOF protocols on \(N-1\) parties into a function that is hard for NOF protocols on \(N\) parties. By iterating this hardness amplification down to just \(2\) parties, \emph{we obtain a reduction that can convert any function that is hard for two-party communication protocols into a function that is hard for NOF multiparty communication protocols.} We find this reduction to be surprising, and believe it may be of independent interest. In more detail, we prove the following.
\begin{lemma}[A recipe for weak NMEs]\label{lem:main:recipe-for-wNMEs}
    Let \(\cond:\zo^n\times\zo^n\to\zo^{r}\) be a two-source condenser for input entropy \(k_0\), output entropy \(k_1\), and error \(\eps_1\). Let \(\NOF:(\zo^{r})^{N-1}\to\zo^m\) be a function with \(\eps_2\)-average-case NOF communication complexity \(>\mu_2:=2^N m\) against non-adaptive number-on-forehead protocols. Then the function \(\nmExt:(\zo^n)^N \to\zo^m\) defined as
    \begin{align*}
\nmExt(x_1,\dots,x_N):=\NOF(\cond(x_1,x_N),\cond(x_2,x_N),\dots,\cond(x_{N-1},x_N))
    \end{align*}
    is a weak non-malleable extractor for entropy \(k=k_0+2r+2\log(1/\eps_1)\) and error \(\eps=2N\eps_1+2^{N(r-k_1)}\eps_2\).
\end{lemma}

Now, in order to actually obtain our explicit weak NMEs, we will just instantiate the above lemma with well-known NOF lower bounds and two-source condensers. It is worth noting that the NOF-hard function is just called over \(N-1\) parties. Thus, in order to construct the NMEs that we actually use for our three-source LREs, we only need an NOF-hard function over \(2\) parties. In other words -- just a function that is hard in the standard two-party communication model (such as any two-source extractor, like the inner product function). We proceed with the proof of the recipe, before instantiating it to get our weak NMEs.

\begin{proof}[Proof of \cref{lem:main:recipe-for-wNMEs}]

Let \(\X_1^0,\dots,\X_N^0,\X_1^1,\dots,\X_N^1\sim\zo^n\) be a collection of independent sources, each with min-entropy at least \(k\), such that \(\X_i^0\equiv\X_i^1\) for all \(i\in[N]\). For convenience, we refer to \(\X_1^0,\dots,\X_N^0\) as the \emph{good sources}, and we refer to \(\X_1^1,\dots,\X_N^1\) as the \emph{shadows}. The goal is to upper bound
\begin{align*}
\Bigg|\nmExt(\X_1^0,\dots,\X_N^0)&\circ\left(\nmExt(\X_1^{b_1},\dots,\X_N^{b_N})\right)_{b\neq\vec{0}\in\zo^N}\\-\U_m&\circ\left(\nmExt(\X_1^{b_1},\dots,\X_N^{b_N})\right)_{b\neq\vec{0}\in\zo^N}\Bigg|.
\end{align*}
To do so, let's start by making this expression less cumbersome. Towards this end, for each \(b\in\zo^N\) we define a random variable
\[
\Z^b:=\nmExt(\X_1^{b_1},\dots,\X_N^{b_N}),
\]
so that we may rewrite the original expression as
\begin{align}\label{eq:recipe-main-equation}
\left|\nmExt(\X_1^0,\dots,\X_N^0)\circ\left(\Z^b\right)_{b\neq\vec{0}\in\zo^N}-\U_m\circ\left(\Z^b\right)_{b\neq\vec{0}\in\zo^N}\right|.
\end{align}
Now, in order to upper bound \cref{eq:recipe-main-equation}, our analysis will proceed in four phases, in which we gradually fix randomness until a strong bound comes for free (from the ingredients that make up \(\nmExt\)).

\paragraph{Phase I: Fixing the shadows.}

The first step is to fix the shadows \(\X_1^1,\dots,\X_N^1\). As these are independent of the good sources \(\X_1^0,\dots,\X_N^0\), this fixing does not affect the latter, and can only make our expression simpler. More formally, using standard fixing arguments, we know there exist fixed strings \(x_1^1,\dots,x_N^1\in\zo^n\) such that if we define \(\tilde{\Z}^b:=(\Z^b\mid\X_1^1=x_1^1,\dots,\X_N^1=x_N^1)\) for each \(b\in\zo^N\), we can upper bound \cref{eq:recipe-main-equation} by
\begin{align}\label{eq:recipe-second-equation}
\left|\nmExt(\X_1^0,\dots,\X_N^0)\circ\left(\tilde{\Z}^b\right)_{b\neq\vec{0}\in\zo^N}-\U_m\circ\left(\tilde{\Z}^b\right)_{b\neq\vec{0}\in\zo^N}\right|.
\end{align}
As a sanity check, note that \(\tilde{\Z}^{\vec{1}}\) is now a fixed constant \(z\in\zo^m\).

\paragraph{Phase II: Fixing the \(1\)-local condenser calls.}

Now, recalling the definition of \(\nmExt\) from the current lemma statement, the next step is to fix \(\cond(\X_1^0,x_N^1),\cond(\X_2^0,x_N^1),\dots,\cond(\X_{N-1}^0,x_N^1)\). Since each output is a deterministic function of a single good source, this will not introduce any correlations. Furthermore, since each condenser doesn't output too many bits, it will decrease the entropy of each good source by just a little. More formally, by combining standard fixing arguments with the min-entropy chain rule (\cref{lem:chain-rule}) and a simple union bound, the following holds. There exist fixed strings \(y_1,\dots,y_{N-1}\in\zo^r\) such that if we define (for each \(b\in\zo^N\) and \(i\in[N]\)) the random variables
\begin{align*}
\tilde{\tilde{\Z}}^b&:=\Big(\tilde{\Z}^b\mid\cond(\X_1^0,x_N^1)=y_1,\cond(\X_2^0,x_N^1)=y_2,\dots,\cond(\X_{N-1}^0,x_N^1)=y_{N-1}\Big),\\
\tilde{\tilde{\X}}_i^0&:=\Big(\X_i^0\mid\cond(\X_1^0,x_N^1)=y_1,\cond(\X_2^0,x_N^1)=y_2,\dots,\cond(\X_{N-1}^0,x_N^1)=y_{N-1}\Big),
\end{align*}
then we can upper bound \cref{eq:recipe-second-equation} by
\begin{align}\label{eq:recipe-third-equation}
\Bigg|\nmExt(\tilde{\tilde{\X}}_1^0,\dots,\tilde{\tilde{\X}}_N^0)\circ\left(\tilde{\tilde{\Z}}^b\right)_{b\neq\vec{0}\in\zo^N}-\U_m\circ\left(\tilde{\tilde{\Z}}^b\right)_{b\neq\vec{0}\in\zo^N}\Bigg|+N\delta,
\end{align}
where each \(\tilde{\tilde{\X}}_i^0\) is independent and has min-entropy at least \(k-r-\log(1/\delta)\). Now, notice that for every \(b\in\zo^N\) with \(b_N=1\), it actually holds that \(\tilde{\tilde{\Z}}^b\) has been fixed to some constant \(z^b\in\zo^m\). Indeed, unwrapping the definition \(\tilde{\tilde{\Z}}^b\) and then \(\nmExt\), observe that every such \(\tilde{\tilde{\Z}}^b\) is actually just a deterministic function of two-source condenser calls of the form \(\cond(\X_i^0,x_N^1)\) (which have been fixed to constants) or \(\cond(x_i^1,x_N^1)\) (which are clearly constant). Thus \cref{eq:recipe-third-equation} can be rewritten as
\begin{align*}
\Bigg|\nmExt(\tilde{\tilde{\X}}_1^0,\dots,\tilde{\tilde{\X}}_N^0)&\circ\left(\tilde{\tilde{\Z}}^b\right)_{\substack{b\in\zo^N:b\neq\vec{0},b_N=0}}\circ\left(z^b\right)_{\substack{b\in\zo^N:b\neq\vec{0},b_N=1}}\\-\U_m&\circ\left(\tilde{\tilde{\Z}}^b\right)_{\substack{b\in\zo^N:b\neq\vec{0},b_N=0}}\circ\left(z^b\right)_{\substack{b\in\zo^N:b\neq\vec{0},b_N=1}}\Bigg|+N\delta.
\end{align*}
Applying the data-processing inequality (\cref{lem:data-processing-inequality}), this is at most
\begin{align}\label{eq:recipe-fourth-equation}
\Bigg|\nmExt(\tilde{\tilde{\X}}_1^0,\dots,\tilde{\tilde{\X}}_N^0)\circ\left(\tilde{\tilde{\Z}}^b\right)_{\substack{b\in\zo^N:b\neq\vec{0},b_N=0}}-\U_m\circ\left(\tilde{\tilde{\Z}}^b\right)_{\substack{b\in\zo^N:b\neq\vec{0},b_N=0}}\Bigg|+N\delta.
\end{align}
Now, let's analyze the remaining \(\tilde{\tilde{\Z}}^b\). Notice that since \(b\in\zo^N\) satisfies both \(b\neq\vec{0}\) and \(b_N=0\), there must be some \(i\in[N-1]\) such that \(b_i\neq 0\). Since we have fixed all the shadows, this means that \(\tilde{\tilde{\Z}}^b\) is a deterministic function of \(\tilde{\tilde{\X}}^0_{-i}\). To make this more formal, we partition the remaining \(b\in\zo^N\) into buckets according to which coordinate is nonzero. In particular, define for each \(i\in[N-1]\) the set
\[
B_i:=\{b\in\zo^n : b_i=1,b_N=0\}.
\]
To make this into an actual partition, simply define \(B_i^\pr:=B_i\setminus(\cup_{h<i}B_h)\) to only include the strings not yet accounted for. Using this partition, we can rewrite \cref{eq:recipe-fourth-equation} as
\begin{align*}
\Bigg|\nmExt(\tilde{\tilde{\X}}_1^0,\dots,\tilde{\tilde{\X}}_N^0)\circ\left(\tilde{\tilde{\Z}}^b\right)_{b\in B_i^\pr,i\in[N-1]}-\U_m\circ\left(\tilde{\tilde{\Z}}^b\right)_{b\in B_i^\pr,i\in[N-1]}\Bigg|+N\delta.
\end{align*}
Furthermore, recall from above that if \(b\in B_i^\pr\), then \(b_i=1\), which means that \(\tilde{\tilde{\Z}}^b\) is a deterministic function of \(\tilde{\tilde{\X}}^0_{-i}\). Moreover, note that each \(\ttZ^b\) is over \(m\) bits, and the total number of elements across all the \(B_i^\pr\) is at most \(2^{N-1}\). Thus there exist deterministic functions, \(g_1,\dots,g_{N-1}\), which output \(2^{N-1}m\) bits in total, such that the above expression is exactly
\begin{align}\label{eq:recipe-fifth-equation}
\Bigg|\nmExt(\tilde{\tilde{\X}}_1^0,\dots,\tilde{\tilde{\X}}_N^0)\circ\left(g_i\left(\tilde{\tilde{\X}}^0_{-i}\right)\right)_{i\in[N-1]}-\U_m\circ\left(g_i\left(\tilde{\tilde{\X}}^0_{-i}\right)\right)_{i\in[N-1]}\Bigg|+N\delta.
\end{align}
Finally, we are ready to proceed to the next phase.

\paragraph{Phase III: Boosting the min-entropy.}

In this phase, our goal is to execute the two-source condenser calls baked into \(\nmExt\) in order to boost the min-entropy of the sources \(\tilde{\tilde{\X}}_i^0\), and prepare them for the final phase, in which we call an NOF-hard function that expects to receive uniformly random input. Towards this end, define for each \(i\in[N-1]\) the random variable \[
\Y_i:=\cond(\tilde{\tilde{\X}}^0_i,\tilde{\tilde{\X}}^0_N),
\]
so that we may rewrite \cref{eq:recipe-fifth-equation} as
\begin{align}\label{eq:recipe-sixth-equation}
\Bigg|\NOF(\Y_1,\dots,\Y_{N-1})\circ\left(g_i\left(\tilde{\tilde{\X}}^0_{-i}\right)\right)_{i\in[N-1]}-\U_m\circ\left(g_i\left(\tilde{\tilde{\X}}^0_{-i}\right)\right)_{i\in[N-1]}\Bigg|+N\delta.
\end{align}
Then, we recall that since \(\cond:\zo^n\times\zo^n\to\zo^r\) is a two-source condenser for input entropy \(k_0\), output entropy \(k_1\), and error \(\eps_1\), it is automatically a \emph{strong} two-source condenser for input entropy \(k_0+r+\log(1/\eps_1)\), output entropy \(k_1\), and error \(\eps_1\) (\cref{fact:every-condenser-is-strong}). Thus, by standard fixing arguments, the definition of strong condenser (\cref{def:condenser}), and a simple union bound, the following holds. We may fix \(\ttX_N^0\) to a string \(x_N^0\in\zo^n\) such that if we define \(\tilde{\Y}_i:=\cond(\tilde{\tilde{\X}}_i^0,x_N^0)\) for each \(i\in[N-1]\), then we may upper bound \cref{eq:recipe-sixth-equation} by
\begin{align}\label{eq:recipe-seventh-equation}
\Bigg|\NOF(\tY_1,\dots,\tY_{N-1})\circ\left(g_i\left(\tilde{\tilde{\X}}^0_{-i}\right)\right)_{i\in[N-1]}-\U_m\circ\left(g_i\left(\tilde{\tilde{\X}}^0_{-i}\right)\right)_{i\in[N-1]}\Bigg|+N\delta+(N-1)\eps_1,
\end{align}
where each \(g_i\) is now a deterministic function of \(\ttX^0_1,\dots,\ttX^0_{i-1},\) \(\ttX^0_{i+1},\dots,\ttX^0_{N-1}\), all \(\{\ttX_j^0\}\) are still independent, each \(\tY_j\) is a deterministic function of \(\ttX_j^0\),  and each \(\tY_j\) has min-entropy at least \(k_1\). Now, note that since \(\tY_j\) is a deterministic function of \(\ttX_j^0\), we may write \(\tY_j=f_j(\ttX_j^0)\) for some deterministic function \(f_j\). On the other hand, by the dependency reversal lemma (\cref{lem:dependency-reversal}), we know that we can furthermore find a deterministic function \(h_j\) and a new random variable \(\A_j\) (which is completely independent of all random variables seen so far) such that \(\ttX_j^0=h_j(f_j(\ttX_j^0),\A_j)=h_j(\tY_j,\A_j)\). This means that any fixing of \(\A_j\) does not affect the distribution or independence of the \(\tY_j\)'s, and furthermore every such fixing turns \(\ttX_j^0\) into a deterministic function of \(\Y_j\). As a result, standard fixing arguments tell us that there exists \emph{some} fixings of \(\A_1,\dots,\A_{N-1}\) such that \cref{eq:recipe-seventh-equation} can be upper bounded by
\begin{align}\label{eq:penultimate-here}
\Bigg|\NOF(\tY_1,\dots,\tY_{N-1})\circ\left(g_i\left(\tY_{-i}\right)\right)_{i\in[N-1]}-\U_m\circ\left(g_i\left(\tY_{-i}\right)\right)_{i\in[N-1]}\Bigg|+N\delta+(N-1)\eps_1,
\end{align}
where each \(g_i\) is now a deterministic function of \(\tY_{-i}\), all the \(\tY_i\)'s are still independent, and all the \(\tY_i\)'s still have min-entropy at least \(k_1\).

\paragraph{Phase IV: Executing the NOF-hard function.}

Finally, notice that the sequence of functions \((g_i)_i\) is exactly a (non-adaptive) number-on-forehead protocol over the remaining random variables \(\tY_1,\dots,\tY_{N-1}\) \(\sim\zo^r\), and each of these remaining random variables has min-entropy at least \(k_1\). Moreover, recall that the functions \(g_1,\dots,g_{N-1}\) output \(2^{N-1}m\) bits in total. Additionally, the lemma statement asserted that \(\NOF\) has \(\eps_2\)-average-case NOF communication complexity \(>\mu_2:=2^N m\) against number-on-forehead protocols. Since \(2^{N-1}m\leq2^N m-2\), \cref{lem:NOF-with-missing-entropy} immediately tells us that \cref{eq:penultimate-here} is at most
\[
\eps_2\cdot2^{N(r-k_1)}+N\delta+(N-1)\eps_1.
\]
Finally, recall that for the condenser to actually work (and be strong), we needed to feed it sources with min-entropy \(k_0+r+\log(1/\eps_1)\). And furthermore, recall that the sources we fed into it had min-entropy at least \(k-r-\log(1/\delta)\). Thus we need to have \(k-r-\log(1/\delta)\geq k_0+r+\log(1/\eps_1)\). In other words, if we set \(\delta=\eps_1\) and start with min-entropy at least \(k\geq k_0 + 2r+2\log(1/\eps_1)\), \cref{eq:penultimate-here} is bounded above by
\[
\eps_2\cdot2^{N(r-k_1)}+N\eps_1+(N-1)\eps_1\leq\eps_2\cdot2^{N(r-k_1)}+2N\eps_1,
\]
as desired.
\end{proof}

Now that we have completed the recipe, let's instantiate it to obtain our explicit weak NMEs (which, in the next section, we'll use to get our LREs). Recall that the weak NME is built from an NOF-hard function and a two-source condenser. We import below the ones that we use.

\begin{theorem}[Finite field multiplication is NOF-hard \cite{ford2013hadamard}]\label{theorem:ffm-is-cylinder-ext}

There exists a universal constant \(c>0\) such that the following holds. Let \(f:(\zo^n)^N\to\zo^m\) be the function \(f(x_1,\dots,x_N):=(x_1\cdot x_2 \dotsm x_N)_{\leq m}\) that multiplies its inputs over \(\F_{2^n}\), and outputs the first \(m\) bits. As long as \(m\leq cn/2^N\), it holds that \(f\) has \(2^{-cn/2^N}\)-average-case NOF communication complexity \(>\mu:=\lceil cn/2^N\rceil\).\footnote{In the original paper \cite{ford2013hadamard}, this result is only explicitly stated for \(m=1\). However, it is straightforward to extend to larger \(m\) - for a formal presentation, see \cite{ourLREs}.}
\end{theorem}

\begin{theorem}[Low-error two-source condenser \protect{\cite[Theorem 31]{ben2019two}}]\label{theorem:low-error-condenser}
For every small enough constant \(\gamma>0\), there exists a constant \(C>0\) such that the following holds. For every \(n\geq k\geq\log^Cn\), there exists an explicit two-source condenser \(\cond:\zo^n\times\zo^n\to\zo^m\) for min-entropy \(k\) with output length \(m\geq k-6k^\gamma\), output entropy \(\ell\geq m-k^\gamma\), and error \(\eps=2^{-k^\gamma}\).
\end{theorem}

Now, let's obtain our explicit weak NMEs.

\begin{proof}[Proof of \cref{lem:main:explicit-wNMEs}]
Simply plug the two-source condenser from \cref{theorem:low-error-condenser} and the NOF-hard function from \cref{theorem:ffm-is-cylinder-ext} into the recipe for constructing weak NMEs (\cref{lem:main:recipe-for-wNMEs}), and set constants \(C,\gamma\) appropriately.
\end{proof}

\dobib

\section{Putting everything together}\label{sec:wrapping-up}

Now that we have our reduction from LREs against NOFs to weak NMEs, and our explicit construction of weak NMEs, we are ready to construct our LREs against NOFs.

\begin{theorem}[LREs against NOFs - \cref{thm:main:LREs}, general version]\label{thm:technical:LREs}
    There exist universal constants \(C,\gamma>0\) such that the following holds. There exists an explicit leakage-resilient extractor \(\Ext:(\zo^n)^N\to\zo^m\) for min-entropy \(k\geq\log^Cn\) against number-on-forehead-protocols with \(\mu=\lceil k^\gamma/2^N\rceil\) bits of communication, which has output length \(m=\lceil k^\gamma/2^N\rceil\) and error \(\eps=2^{-\lceil k^\gamma/2^N\rceil}\).
\end{theorem}
\begin{proof}
Simply combine \cref{lem:main:reduction,lem:main:explicit-wNMEs}, resetting constants \(C,\gamma\) appropriately.
\end{proof}

Setting $N=3$ in \cref{thm:technical:LREs}, we obtain the following LRE against NOF protocols.

\begin{corollary}[LREs against NOFs - \cref{thm:main:LREs}, restated]\label{thm:main:LREs:restated}
There exist universal constants \(C,\gamma>0\) such that the following holds. There exists an explicit three-source extractor \(\Ext:(\zo^n)^3\to\zo^m\) for min-entropy \(k\geq\log^Cn\), which is leakage-resilient against number-on-forehead protocols with \(\mu=\lceil k^\gamma\rceil\) bits of communication, and which has output length \(m=\lceil k^\gamma\rceil\) and error \(\eps=2^{-\lceil k^\gamma\rceil}\).
\end{corollary}

Given this LRE against NOFs, it is now straightforward to obtain our extractors for adversarial sources.

\begin{theorem}[Extractors for adversarial sources - \cref{thm:main:extractors-adversarial-sources}, restated]
There exist universal constants \(C,\gamma>0\) such that the following holds. There exists an explicit extractor \(\Ext:(\zo^n)^N\to\zo^m\) for \((N,K,n,k)\)-adversarial sources with \(K=3\) good sources of min-entropy \(k\geq\log^Cn\), which has output length \(m=\lceil k^\gamma\rceil\) and error \(\eps=2^{-\lceil k^\gamma\rceil}\).
\end{theorem}
\begin{proof}

Let \(\mathsf{LRE}:(\zo^n)^3\to\zo^m\) be the LRE against NOFs from \cref{thm:main:LREs:restated}, and define the function \(\Ext:(\zo^n)^N\to\zo^m\) as
\[
\Ext(\X):=\bigoplus_{1\leq a<b<c\leq N}\mathsf{LRE}(\X_a,\X_b,\X_c).
\]
Without loss of generality, we may assume that the good sources in \(\X\) are \(\X_1,\X_2,\X_3\). Fix all \(\X_i,i>3\), and observe that \(\Ext\) becomes
\[
\Ext(\X)=\LRE(\X_1,\X_2,\X_3)\oplus f_1(\X_2,\X_3)\oplus f_2(\X_1,\X_3)\oplus f_3(\X_1,\X_2)
\]
for some deterministic functions \(f_1,f_2,f_3\). By combining the definition of LREs against NOFs with a standard application of the data-processing inequality, it follows that \(\Ext(\X)\) is close to uniform.
\end{proof}

\dobib

\bibliographystyle{alpha}
\bibliography{references}

\end{document}